\documentstyle[11pt]{article}
\parskip=\medskipamount
\begin{document}

\begin{LARGE}
Microscopic model for spreading of a  
two-dimensional monolayer
\end{LARGE}

\vspace{1cm}

\begin{Large}
G.Oshanin$^{a,b}$, J.De Coninck$^a$, A.M.Cazabat$^c$ 
and M.Moreau$^b$\\
\end{Large}

\vspace{0.5cm}

$^a$ Centre de 
Recherche en Mod\'elisation Mol\'eculaire, 
Universit\'e de Mons-Hainaut, 20 Place
du Parc, 7000 Mons, Belgium\\

$^b$ Laboratoire de Physique Th\'eorique des Liquides,
Universit\'e Paris VI, 4 Place Jussieu, 
75252 Paris Cedex 05, France\\

$^c$ Laboratoire de Physique de 
la Mati\`ere Condens\'ee,
Coll\`ege de France, 11
Place M.Berthelot, 75231 Paris Cedex 05, France\\

\vspace{0.5cm}

We study the behavior  
of a monolayer, 
 which occupies initially
a bounded region on an ideal crystalline surface
and then evolves in time  due to
random hopping 
of the monolayer particles. 
In the case when the initially occupied region 
is the half-plane $X \leq 0$, we 
determine explicitly, in terms of an analytically solvable 
mean-field-type approximation,  
the mean displacement $X(t)$
 of the monolayer edge.  
We find that $X(t) \approx A \sqrt{D_{0} t}$, 
in which law 
 $D_{0}$ denotes the bare diffusion coefficient and 
the prefactor $A$ is a  function of
the temperature and of the particle-particle 
interactions parameters.  
We show that  $A$ 
can be greater, equal or less than zero, 
and  
specify the critical parameter which 
distinguishes between the regimes
 of spreading
($A > 0)$, partial wetting ($A = 0$)
 and dewetting ($A < 0$).

\section{Introduction}

It has been well appreciated since the 
pioneering work by Hardy
\cite{hardy} 
that  
a liquid droplet spreading on a solid, although it undergoes
no visible change in shape, emits 
 a very thin invisible film - 
the precursor, which advances 
at a seemingly faster rate than the nominal
 contact line. 
Hardy 
was able to detect its presence by observing a significant change
in the value of the static friction of the surface.
Stating that he was unable to conceive of a 
mechanism by which the film can be emitted and spread further along the solid
substrate, Hardy proposed that spreading  of the film
 occurs by a process involving 
a continual condensation of vapor. Bangham and Saweris  \cite{bang}
have demonstrated, however, that 
such a 
film shows up even in the 
absence of any vapor fraction, suggesting thus
that the physical mechanism giving rise 
to the precursor film 
can be 
different
from the evaporation/condensation  scheme. 
Later, using ellipsometric and interferometric
techniques, Bascom et al. \cite{bas}
investigated the spreading of the precursor film from a more quantitative point of view.
They examined the behavior of various non-polar liquids on clean metal 
surfaces in the presence of both
saturated and unsaturated air and 
concluded that the precursor film is always present;
making the air saturated or unsaturated with vapor, roughening the surface and purifying the liquids
does
not eliminate the film, but only affects the speed 
at which it spreads over the solid. 
The thickness of the film was also found to depend essentially 
on the liquid/solid system under study; it can
be as small as molecular size (several angstroms) or 
can amount to hundreds of angstroms (see
\cite{caz0} for a review).  Particularly,  the droplet of squalane spreading
on stainless steel exhibited a precursor with a thickness of approximately $20$ angstroms.

Spreading rates and dynamical shapes 
of advancing precursor films have been 
studied thoroughly for many 
years, both experimentally and
theoretically (see \cite{bas,caz0,pgg0,der0,lyk,caze} and 
references therein).  These studies have resulted in 
a rather good understanding of the problem. 
It was realized 
that the precursor film appears not because 
of the condensation of the vapour fraction, which process may, 
of course, exist but plays a minor role.
Rather, such a film  
extracts from the  droplet and advances along the solid surface
due to the  presence of attractive 
interactions between the fluid molecules and the solid
atoms \cite{bas,caz0,pgg0,der0,lyk,caze}. 
Later, it even became possible
to elaborate consistent hydrodynamic theory of spreading of 
thin precursor films, based on the celebrated 
lubrication approximation of fluid
mechanics 
\cite{caz0,pgg0,der0,joa}. 
In particular, this considerable theoretical advancement 
 allowed 
to resolve an old-standing enigma 
concerning the dynamics of the nominal contact line: 
it was well documented 
experimentally (see, e.g. \cite{pgg0} and references therein) that 
in the complete spreading regime 
the radius $R$ of the nominal contact line shows a slow growth with 
time, $R \sim t^{1/10}$, where the exponent $1/10$ is universal, i.e.
dependent only on the geometry and 
independent of the precise nature of the liquid/solid system.  
As a matter of fact, in this law even the prefactor appears to be
insensitive to the spreading power $S$, which is the characteristic of
 a given liquid/solid
system and equals the difference of the surface tensions of the solid-vapour,
solid-liquid and liquid-vapour interfaces respectively. 
On the other hand,
conventional analysis, which considers the gradient  of the liquid/solid 
free energy as
the driving force of spreading,  
predicts much faster growth which is, moreover,
dependent on the spreading power. The hydrodynamic picture developed in 
\cite{pgg0,joa} has found the solution of this controversy, showing
that the  spreading power $S$ is totally dissipated into the precursor
film, such that the dynamics of the nominal
contact line is actually not affected by $S$. Besides, it has been recognized
that contrary to 
the general belief, the final state
in complete ($S \geq 0$) spreading of a liquid droplet is not necessarily a 
monolayer covering the solid surface: it is only 
the  macroscopic part of the droplet
which spreads completely;
the precursor film, however, can cease to spread further 
after the macroscopic part
of the droplet is exhausted and form 
a stable, nearly planar droplet-like structure
- the so-called "pancake" \cite{pgg0,joa,joab,ruck1,ruck2}.  The thickness of the
pancake is fixed by a competition between $S$ and
 long-range attractive interactions
with the substrate, which tend to thicken the film. For pure van der Waals
interactions the thickness $e_{p}(S)$ of the pancake is $e_{p}(S) \approx
 a (3 \gamma/2
S)^{1/2}$, where $a$ is the molecular size and $\gamma$ is the macroscopic surface
tension of the liquid. In general, $\gamma/S \sim 1$ and $e_{p}(S) \sim a$, but if
$S$ is small, $e_{p}(S)$ can be relatively large.
 Therefore, within the framework of the hydrodynamical
approach it is  the pancake thickness 
$e_{p}(S)$ which sets the upper bound on the thickness of the precursor film. 
Details concerning 
different "pancake" structures and their experimental observations can be found in a 
recent review \cite{cazy}.

\subsection{Experimental studies of spreading of molecularly thin films.}

Hydrodynamic approach, however, presumes a certain lower cut-off length
comparable to the molecular size, 
below which
it is not justified. With the advent of modern experimental techniques, 
capable of
studying behavior of precursor films whose thickness is in the molecular 
range, it has become
clear that the developed theoretical concepts concerning the dynamics and
equilibrium
properties of thin wetting films can only aplly to sufficiently 
thick films. In fact,  for molecular
films significant departures from the hydrodynamic picture have been 
observed experimentally
\cite{caza,cazb,cazc,cazd,cazf}. 

Extensive ellipsometric measurements \cite{caza,cazb,cazc,cazd,cazf}
 have elucidated several remarkable features, which seem to be quite
 generic for 
spreading molecularly thin films:

(i) Experimental measurements 
carried out on different substrates and
 with different types of simple liquids,
polymeric and surfactant melts, showed unambiguously  
that the 
 radius of the precursor film
grows with time $t$ as $t^{1/2}$. 
The exponent $1/2$  appears
to be completely independent 
of the nature of the species
involved. The latter affects  the spreading    
rate only through the prefactor in the
$\sqrt{t}$-law.  Essentially the same behavior 
has been discovered 
in the capillary rise geometry, 
in which
a vertical solid wall is put into contact
with a bath of liquid. The length
of the molecularly
thin film extracting from meniscus
 and climbing upwards along the wall was also shown to grow 
in proportion to the square-root of time
 \cite{cazd}. This shows that the $\sqrt{t}$-law is seemingly independent of
the precise curvature of the film's edge, which is a circular line
in case of the sessile drops and planar in the capillary rise geometries.\\

(ii) It was found 
that the particle density 
along the film is not necessarily constant. In several experimental situations,
particularly, for the droplets of squalane, 
the radial density was seen to
vary strongly
 with the distance from the nominal contact line and  
the variation was progressively  more pronounced for larger 
 spreading rates. \\

(iii) For intermediate-energy substrates a fascinating transient regime
of "terraced spreading" was discovered   \cite{caza,cazb}, in
 which several molecularly thin precursor films
extract from the macroscopic droplet and 
spread on top of one another each one growing
as $\sqrt{t}$.  
Such a regime may last within a considerably 
long period of time, until the layers on the top run out
and one eventually ends up with a bounded monolayer
on the solid surface. Depending on the physical conditions, 
the monolayer may then 
either continue to spread
with the radius growing in proportion to $\sqrt{t}$, 
(but with a different prefactor, of course), 
or may remain in form of a wetted spot \cite{cazd}.
The remarkable "terraced wetting" effect has been  also 
seen in the capillary rise geometries. Besides, recent
numerical, Molecular Dynamics simulations of liquid drop spreading were
able to reproduce both the "terraced" wetting spreading regime 
and the $\sqrt{t}$-law 
 (see \cite{ort,col,ala} and references therein).
 
\subsection{Theoretical studies of spreading of molecularly thin films.}

Theoretical analysis of the physical mechanisms underlying
the seemingly universal $\sqrt{t}$-law
and the "terraced wetting" phenomenon followed
three different lines of thought:\\

In \cite{pgg2} an analytical 
description of the "terraced wetting"
phenomenon has been proposed, in which the
liquid drop on a solid surface 
was considered as a completely
 layered structure, 
each layer 
being a two-dimensional (2D) incompressible 
fluid  of molecular thickness but with a macroscopic 
radial extension.
The interaction energy of a molecule in the $n$th 
layer with the solid substrate
was taken in the general form as a negative, decreasing function of the distance from the substrate. 
Next, it was supposed that spreading proceeds
by filling the  successive layers by the molecules 
from the above layers, which process is
favored by the attractive interactions with  the solid. 
In each layer there is
a horizontal, radial particle current and vertical 
permeation fluxes, one from the upper layer
and one towards the lower layer, which appear  
to be located in a very thin "permeation ribbon" 
just at the droplet surface - the core
of the droplet seems to be a "stagnant" liquid with 
respect to the vertical mass transfer.
In such an approach, 
 de Gennes and Cazabat \cite{pgg2} 
have found that  whenever the distinct layers grow at a comparable rate, they
 grow  in
proportion to $t^{1/2}$. In case when the precursor film grows at a much faster rate than
all other layers, the model predicts that the precursor grows in proportion to 
$(t/ln(t))^{1/2}$.  
 Apart from the theoretical prediction of the  $\sqrt{t}$-law,
the model allows to 
draw  certain 
conclusions concerning 
the appearence of the
"terraced wetting"  regime. To some extent, the underlying assumptions
and predictions
of this model concerning the intermediate time behavior
  were confirmed by Molecular Dynamics
simulations \cite{vil}.  

An alternative description within the framework  
of non-equilibrium statistical 
mechanics has been elaborated in \cite{joela,joelb,joelc}. 
Here, an interfacial 
model for the non-volatile
fluid edge has been developed and analysed 
in terms of Langevin dynamics for the displacements of
horizontal solid-on-solid (SOS) strings $\{h_{j}\}$ at
increasing heights $j = 1, ... , L$ from the substrate, which strings 
have essentially the same meaning as the layers in the de Gennes-Cazabat model
\cite{pgg2};  $h_{j}$ can be thought of as the radius of the $j$th layer.
Anticipating
the discussion of results obtained in the present paper, we 
will describe this promising theoretical approach giving some more
details.
In the model developed in \cite{joela,joelb,joelc} the energy
$U(\{h_{j}\})$ of a given configuration $\{h_{j}\}$ was described by
\begin{equation}
U(h_{0}, h_{1}, ... , h_{L}) \; = \; \sum_{j = 1}^{L} P(h_{j} - h_{j-1}) \; - 
\; \mu_{0} \; h_{0},
\end{equation}
where $h_{0}$ is the length of the "precursor" film, $\mu_{0}$  
is the wall tension and
the function $P(h_{j} - h_{j-1})$ 
determines  the contribution due to the surface energy.
Explicitly, it was taken as
\begin{equation}
P(h_{j} - h_{j-1}) \; = \; J \; \sqrt{1 \; + \; (h_{j} - h_{j-1})^{2}},
\end{equation}
where the parameter $J$ is related to the surface tension; 
thus, the spreading power $S$ for such a model
is given by $S = \mu_{0} - J$.  In the limit $h_{j} - h_{j-1}  \ll
 1$, the function in Eq.(2) reduces to the standard Gaussian form, i.e., 
\begin{equation}
P(h_{j} - h_{j-1}) \; \approx \; J \; (h_{j} - h_{j-1})^{2},
\end{equation}
while for  $|h_{j} - h_{j-1}|  \gg 1$ it obeys
\begin{equation}
P(h_{j} - h_{j-1}) \; \approx \; J \; |h_{j} - h_{j-1}|
\end{equation}
In other words, Eq.(2), which bears a certain relationship with the Lifschitz
equation describing the time evolution of an interface due to the effects 
of the surface tension \cite{lif}, supposes the following, quite realistic behavior of the
interfacial energy:
for small distortions of the interface the surface energy is Gaussian, 
while the cost of 
a large distortion is linear with the
distortion size. 
Dynamics of the strings $\{h_{j}\}$ is then described by the set of $L$ coupled
Langevin equations
\begin{equation}
\xi \; \frac{\partial h_{k}}{\partial t} \; = \; - \; \frac{\partial U(\{h_{j}\})}{\partial h_{k}} \; + \;
f(h_{k};t),
\end{equation}
where $\xi$ denotes the friction coefficient, which is supposed to be the same for all layers,
 and $f(h_{k};t)$ 
is Gaussian, delta-correlated noise.
The model in Eqs.(1),(2) and (5) allows for an analytical, 
although rather complicated solution, 
which shows an extraction of the precursor
film and "terraced" forms of the dynamical
 thickness profiles. It predicts that for $\mu_{0} > J$, ($S > 0$), and 
for sufficiently short precursors, (such that the 
dominant contribution to the surface 
energy 
is given by Eq.(3)), the length of the film increases with time as
\begin{equation}
h_{0}(t) \; \sim \; \sqrt{(\mu_{0} - J) \; t},
\end{equation}
which resembles the experimentally observed behavior. For sufficiently large
precursor, for which the energy increases linearly with the length, Eq.(4), 
the  layer on top of substrate is found to show a faster growth
\begin{equation}
h_{0}(t) \; \sim \; (\mu_{0} - J) \; t 
\end{equation}
Finally, it was found that exactly 
at the wetting transition point $\mu_{0}  =  J$, the precursor film 
advances in proportion to 
\begin{equation}
h_{0}(t) \; \sim \; \sqrt{t \; \ln(t)} 
\end{equation}
Consequently, this model predicts  that at very 
large times
the advancing
 precursor film attains  
a constant velocity; the $\sqrt{t}$-law is thus found 
only as a transient stage. Besides, dynamics of the layers
at large distances from the substrate and, respectively, of the macroscopical
dynamical  contact
angle disagree with experimental data.
Apparently, this inconsistency with experiments can be 
traced back to the fact that
focusing on the evolution of the interface only, 
the model neglects dynamics in the liquid phase and 
thus discards the energy losses due to viscous flow pattern, 
generated in the spreading
droplet. In \cite{joela,joelb,joelc} a viscous-type dissipation is assumed 
with constant friction coefficient,
which represents rather generic and oftenly loose 
assumption used
in the descriptions of the phase-separating boundary dynamics in terms of the
time-dependent Landau-Ginzburg-type model. In fact, the model 
underestimates the dissipation
 in each layer $h_{j}$. 
Account of the energy losses in viscous flows in the core region of the droplet
and of the  
dissipation in the vicinity of the
solid substrate actually results in the overall 
viscous-type dissipation, which is not 
surprising for the system with many degrees of freedom. 
The friction coefficient, however,
turns out to be
weakly dependent on the height above the substrate and, what is 
essentially
more important, appears to be an increasing function of $h_{j}$.
Therefore, these dissipation chanels should be certainly taken into account
within the framework of  
the powerful theoretical
approach proposed in \cite{joela,joelb,joelc}, which may result
in a consistent 
dynamical theory of partial and complete wetting valid for all scales. 
Such improvements  are currently under investigation \cite{joelx}.

Lastly,  a microscopic dynamical model for 
spreading molecularly thin  films 
has been devised
in \cite{bura,burc}. Here 
the 
film was considered as a two-dimensional
hard-sphere fluid with particle-exchange dynamics. 
Attractive interactions between the particles in the precursor
 film
were not ostensibly included into the model, but introduced 
in a mean-field-type
way - it was supposed that the film is enclosed by 
the SOS-model
interface, in which the parameter $J$ was treated as some 
(not specified in \cite{bura,burc})
 function of the amplitude of the
particle-particle attractions.   
The film
was assumed to be
connected to a reservoir of infinite capacity - the macroscopic drop.
The rate at which the reservoir may add particles into the film was
related to the local particle density in the film in the vicinity of
the nominal contact line and to the strength of
the van der Waals attractive interactions between the fluid
particles and solid atoms  
within the framework of the standard Langmuir adsorption theory.
Contrary to \cite{pgg2} and to the hydrodynamic picture of \cite{pgg0},
the model in  \cite{bura,burc} emphasized the issues of compressibility
and molecular diffusion at the expense of the hydrodynamic flows; 
it was assumed
that the reservoir and the film are in 
equilibrium with each other, so that there is no flow of particles from the reservoir
which pushes particles to move along the substrate away of the droplet.
In this approach, 
the $\sqrt{t}$-law for growth of the 
film was first analytically obtained
for the capillary rise geometries and it was actually found that
the density in the film does varies strongly with the distance from the
reservoir. This agrees, at least qualitatively, with  experimental observations 
(see
the Introduction, 1.1, (ii)), but no direct comparison was made, as yet. Clearly, 
the factor which makes
such a comparison quite awkward is that the spreading rate is 
expressed through the
parameter $J$, which is supposed to be some known parameter. 
As it will be made clear below, this
parameter depends on the particle-particle attractions and
 moreover, on the density
distribution in the precursor film.
Further on,  the 
critical conditions under which 
spreading of the precursor film may take place
were established in \cite{bura,burc}. 
It was also suggested that
the physical mechanism underlying the $\sqrt{t}$-law 
stems from diffusive-type
transport of vacancies from the edge of the advancing film to the
macroscopic liquid edge, where they perturb the equilibrium between
the macroscopic drop and the film and 
get filled with fluid particles from the
macroscopic liquid drop. 
In \cite{burd} this picture was extended to the case
of  sessile drops  and it was shown analytically
that  the curvature corrections result only 
in a weak slowing down of the precursor spreading;
the film radius grows in this case as 
$(t/ln(t))^{1/2}$, 
which prediction agrees with 
\cite{pgg2}.

To close this introductory part of our paper
we mention several analytical studies of the process, 
which can be thought of as the
reverse counterpart of wetting, - 
dewetting of microscopically thin liquid films from solid substrates. 
Recently  Ausserr\'e et al. \cite{pgg1} investigated analytically 
 dewetting
of a monolayer, which was assumed to proceed by nucleation 
of holes (bare regions)
and creation of "towers" - two-layer regions.  
Considering the monolayer
as an incompressible 2D liquid, it was shown that the hole
radius $R$ (or the radius of a "tower") grows with time
in proportion to $(t/ln(t))^{1/2}$.  This means that in the monolayer regime
the dewetting process proceeds essentially slower 
than in the case 
of mesoscopically or macroscopically
thick films, for which the behavior $R \sim t$ is generic \cite{broch,pgg3,red,shana}.
Another interesting example of a (forced) dewetting of a monolayer
 was discussed in
\cite{tos} and concerned with
 the squeezing  of a molecularly thin liquid film out of a narrow gap
between two immobile solids.  In \cite{tos} the
 mechanism responsible for squeezing was attributed to the process of spontaneous
opening of holes in liquid layers;  the holes are subsequently get filled  
by deforming solid material
exerting pressure on the hole boundaries. Viewing solids as isotropic, 
structureless
 elastic media and  the
liquid phase in a 
lubricated contact between
two solids as a sequence of 
layers, each layer being  
an incompressible 2D  liquid,  Persson and Tosatti \cite{tos}
 were able to estimate
the critical radius of the hole, necessary to initiate
 further squeezing, and to 
define the
rate of the removal process after the nucleation of
 a critical hole has occured. 
 It was shown that the radius of the hole, which
 exceeds initially the critical
value, grows in time again in proportion to $(t/ln(t))^{1/2}$.

\section{The objectives and a brief outline of the paper.}

In this paper we study analytically, in terms of a
stochastic microscopic model, the behavior of
 a liquid monolayer 
in a situation, in which a monolayer occupies initially
only some part of solid surface - the half-plane $X \leq 0$, Fig.1,
 and then is allowed to evolve in time due to the
thermally activated random motion of the monolayer particles. 
Here we aim to calculate the 
time dependence of the mean displacement
of the  monolayer edge, 
defined as the position of the rightmost monolayer
particles (Fig.1), to determine the prefactor in this dependence in terms of the
interaction parameters and  
the edge tension of the monolayer.

We note that the model to be considered here clearly
shares common features with many
dewetting and wetting experimental situations and models, which were described in the
Introduction.
A monolayer in such a non-equilibrium 
configuration appears, for instance,  at the 
 late stages of spreading, when 
the liquid drop feeding the precursor film gets exhausted or
 in the situation when the monolayer on the solid surface
is perturbed by a sudden removal
of some amount of particles or by nucleation of a dewetted region - 
a circular hole
or a patch. Such a model  applies, after some minor modifications, to the dynamics of 
 ultrathin liquid columns in nanopores or in  narrow slits between  solid surfaces (Fig.2).
It can also serve as a microscopic description of the process of 
Ostwald ripening of voids, spinodal decomposition or island formation in two-dimensional
adlayers.  We note also that in view of the above-mentioned results concerning spreading dynamics, 
we expect that the precise geometry of the two-phase region is not very important;
the difference between the case when the front of the monolayer  is planar, as we consider here,
and the case when it is a circular closed line, what should be for circular
dewetted holes,  can be 
only in the appearence of logarithmic in time corrections to the $\sqrt{t}$-law, 
important at times when the displacement of the edge becomes comparable to the radius of the edge
curvature.
An important point is that both the region occupied by the monolayer
and the initially dewetted region should be macroscopically large.

As opposed to \cite{pgg2}, \cite{pgg1} 
and \cite{tos}, we will regard  the monolayer as an
essentially discrete,
molecular liquid composed of
interacting particles moving randomly on 
an ideal crystalline surface. Particles migration is assumed
to be activated by the solid atoms vibrations and will be described 
using the standard Kawasaki picture for particles exchange dynamics under
long-range particle-particle interactions.
 The picture we  make use thus follows closely  
the model elaborated   
in \cite{bura,burc}, being 
different from the latter in two important aspects: first, 
the  long-range attractive interactions between the liquid
particles are here explicitly
included into the dynamics, and, consequently, 
our results will be expressed in
 terms of the 
interaction parameters.
Second, the reservoir of particles is absent, 
which allows us to study within a unified approach
dynamics of both spreading and dewetting processes.  
We also hasten to remark that with regard to other 
dynamical wetting theories, this model is  related to
the Molecular Kinetic theory of wetting dynamics, proposed and developed
by Blake et al. 
\cite{blakea,blakeb,mich}. In this theory, which emphasizes
 the dissipation in the vicinity of
the nominal contact line at expense of 
 the dissipation due to viscous flows in the "bulk" droplet,
the analysis of dynamics of spreading 
 liquid droplet 
was reduced to a mean-field-type consideration of the
forced, thermally activated motion of fluid particles which appear
 directly
at the droplet edge. In our case, however, the driving force
is not assumed $\it a \; priori$, 
but is found consistently as the result 
of the cooperative behavior, associated with
the interplay between the long-range attractive particle-particle 
interactions and repulsion at shorter scales. 
As well, we deal here with simultaneous
random motion of all particles in the film, not reducing the problem to
consider the dynamics of only particles at the edge. 

Next, we will make several simplifications, compared to
 real physical systems.  In what follows we will assume that 
creation of "towers" and particles evaporation
 in the direction normal
to the solid surface are completely suppressed
 by the liquid-solid attraction,
and hence we will constrain our consideration to the
 system which  always remains in a two-dimensional world.
Shortcomings of this picture will be discussed below. 
These simplifications  will permit us to
 focus exclusively on the
 dynamical processes which take
place in two-dimensions and thus to single out
 the behavior which
 stems from the interplay
between the compressibility and the 
intermolecular interactions.  We note also 
that these mechanisms  are
 entirely complementary to
those discussed in \cite{pgg2,pgg1,tos} and, consequently,
  understanding
 of their
impact on the monolayer evolution is necessary for 
a complete picture of the phenomenon. 
We remark also that 
such an assumption can be clearly relaxed for liquids in confined geometries, 
e.g. in nanopores or in  
narrow slits between solids, where the geometry
itself rules out an appearance of the vapour phase and thickening of 
liquid films (Fig.2).
Here we will focus, however, solely on the situation with a monolayer on top of open solid
surface; relevant cases, as depicted in Fig.2,  will be discussed elsewhere.

The paper is outlined as follows: In section 3 we 
describe the model and write down basic equations.
In  section 4 we discuss an approximate approach to the solution of 
dynamical equations.
Section 5 presents the results. Finally, in section 6 we conclude with a brief
summary of our results and discussion.

\section{The model and basic equations.}

We proceed further  
with more precise definitions related to the model to be studied here
(see also \cite{bura,burc} for a detailed discussion). 
Particles of the monolayer 
experience two types of 
interactions; 
interactions with the 
solid atoms (SP) 
and mutual interactions
with each other (PP). The SP interactions are characterized  by a 
repulsion at short scales and a
weak attraction at longer distances. 
The SP repulsion  keeps the
monolayer particles at some short distance apart 
from the surface, while 
the attractive part of the SP potential
hinders particles desorption. Following \cite{bura,burc} we assume here
that the SP interactions correspond to the limit of the
 intermediate 
localized adsorption \cite{adam,clark}: the monolayer
 particles are neither completely fixed
in the potential wells created by the SP interactions 
(Fig.1), nor completely
mobile. Potential wells   
are very deep with respect to desorption
(desorption barrier $U_d \gg k T$) so that only a monolayer 
can exist, but have a 
much lower energy barrier $V_l$ 
against the lateral movement
across the surface, $U_d \gg V_l > k T$. In this regime an
 adsorbed  particle 
spends a considerable part of its time at
the bottom of a potential well and  jumps sometimes, due 
to the thermal activation,
from one potential minimum to another. 
Thus, on a macroscopic time scale the particles
do not
possess any velocity.  

We note that such a type of random motion
is essentially different of the standard hydrodynamic 
picture of particles random motion
in the two-dimensional "bulk"
 liquid phase, e.g., in free-standing liquid films,
in which case there is a velocity 
distribution and spatially random motion
results from the mutual
particle-particle interactions; in this case the dynamics  
can be only approximately considered as an activated
 hopping of  particles, confined to some effective 
cells by the 
potential field of their neighbors,
between a lattice-like structure of such cells
(see, e.g. \cite{eyr,dev}).  In contrast to the 
dynamical model to be studied here,
standard two-dimensional hydrodynamics pressumes that the 
particles do not interact with the underlying solid.  
In realistic systems, of course, both the particle-particle scattering 
and scattering by the potential wells
due to the interactions with the host solid, as well as the corresponding dissipation,
 are crucially important \cite{dif,zang};
the latter, particularly, remove the infrared divergencies in the 
dynamic density correlation functions
and thus make the transport coefficients  finite \cite{maz,gor}.  
Complete dynamical description of particles migration on the solid surface
 can be approached apparently 
along the lines proposed 
in \cite{maz,gor} or, on a microscopic level, in terms of the 
cellular automaton-type 
description of \cite{pom}; here we will be thus concerned only 
with a certain approximate
model of particles dynamics, appropriate for situations in which 
the particle-particle interactions are
essentially weaker than the particle-solid interactions.
We note that such an assumption actually makes sense since the latter are
usually at least ten times greater than the PP interactions, and therefore should not be 
appreciably affected by the lateral interactions of adsorbed particles.

Turning next to the particle-particle interactions, we 
suppose that these are additive
and central, i.e. the interaction potential $U(\vec{r}_{j},\vec{r}_{i})$
depends only on the distance $r = |\vec{r}_{j} -  \vec{r}_{i}|$ of
 separation of the $j$th and $i$th
particles, $U(\vec{r}_{j},\vec{r}_{i}) = U(r)$.
 Particles are assumed spherical so that no orientation effects
 enter and we take  
that the potential energy between a pair of adsorbed molecules
 is given by
\begin{equation}
U(r) \; = \; \left\{\begin{array}{lll}
+ \; \infty& \mathrm{for}&r \; < \; \sigma\\
- U_{0}(T) (\sigma/r)^{6}& \mathrm{for}&r \; \geq \; \sigma,\\
\end{array}
\right.
\end{equation}
i.e. we use the "hard-sphere" core and the usual  $r^{-6}$ attractive
 term for large $r$; the minimum
occurs at $r = \sigma$ for which $U(r = \sigma) = - U_{0}(T)$. 
The argument $(T)$ in the parameter $U_{0}(T)$,
$U_{0}(T) \geq 0$, 
 signifies that in   general case, this property can be
 dependent on the temperature. 
Particularly, for the Keesom-van der Waals interactions
one has $U_{0}(T) \sim 1/T$.
 For the London-van der Waals particle-particle
 interactions $U_{0}(T)$ does
not vary with the temperature. As we have already mentioned, we will
 suppose in what follows that 
the amplitude of the particle-particle attraction
 $U_{0}(T)$ is less than the barrier for the lateral
motion, i.e. $V_{l}$, such that the particle-particle
interactions do not perturb significantly the array of
 potential wells created by the particle-solid
interactions.

Now, we specify the particle dynamics more precisely (see also \cite{bura,burc}). 
Under the physical conditions as described above,
we can regard 
the particles dynamics on the solid surface
as
an activated hopping between the 
local minima of an array of 
potential
wells, created due to the SP interactions \cite{adam,clark}.
Thus particles' migration on the surface proceeds by rare 
events of hopping from one well to another in
its neighborhood. The hopping events are separated by the 
time interval $\tau$, which is the time a given
particle typically spends in each well vibrating around its 
minimum; $\tau$ is related to the temperature
$T = \beta^{-1}$, the barrier for lateral motion $V_{l}$ and 
the frequency of solid atoms vibrations
$\omega$ through the Arrhenius formula. 
We thus may estimate the diffusion coefficient for such a 
motion (which will be exactly the diffusion
coefficient of an isolated particle on the solid surface) as 
$D_{0} \approx \l^{2}/z \tau$,
where $z$ is the coordination number of the lattice of wells 
and $l$ is the interwell spacing. In what
follows we will suppose that $l \approx \sigma$, i.e. that  the 
radius of the particle-particle hard-core and the interwell spacing are
approximately the same. 
Now, diffusion coefficient  $D_{0}$ will be the only pertinent 
parameter describing the evolution of the local density in the monolayer
in absence of the PP
interactions. 
When  the latter are present, as we actually suppose here, 
dynamics of any given particle is fairly more
complicated and is coupled to
the instantaneous configuration of the monolayer
 particles (see, e.g. \cite{gom,leb,leba,tur} for
discussion). 
That is, for any particle, 
releasing from the well with radius-vector $\vec{r}$, not
 all hopping directions are equally probable
and the particle has a tendency to 
follow 
the local gradient of the energetic surface
$\rm{U}
(\vec{r};t)$, created by the mutual PP attractions.  
On the other hand,  hard-core repulsion imposes sterical constraints 
preventing particles crossing and thus the double occupancy of any potential well.
 More specifically, we will account for the PP interactions as follows:
we will suppose that releasing  at time moment $t$ 
from the well with radius-vector $\vec{r}$
any given  particle first "choses" the direction
of jump with the (position- and time-dependent) probability 
\begin{equation}
p(\vec{r}|\vec{r'}) 
\; = \; {\rm Z}^{-1} \;   
\exp\left( \frac{\beta}{2} \;
 \left[\rm{U}
(\vec{r};t)  
-  \rm{U}
(\vec{r'};t)\right]\right),  
\end{equation}
where $\vec{r'}$ is the radius-vector of one of $z$ wells 
neighboring to the well at position 
$\vec{r}$, 
${\rm Z}$ is the normalization factor,  defined as
\begin{equation}
\rm Z \; = \; \sum_{\vec{r'}} \exp\left( \frac{\beta}{2} \;
 \left[\rm{U}
(\vec{r};t)  
-  \rm{U}
(\vec{r'};t)\right]\right),  
\end{equation}
in which the sum runs over all wells neighboring to the well as 
position $\vec{r}$, and
the PP energy landscape is determined by 
\begin{equation} 
{\rm U}
(\vec{r};t) \; = \; - \; U_{0}(T) \; \sigma^{6} \; \sum_{\vec{r''}} 
\frac{\eta(\vec{r''};t)}
{|\vec{r} - \vec{r''}|^{6}}
\end{equation}
In the latter equation the summation with respect to $\vec{r''}$ 
extends over the entire surface, excluding $\vec{r''} = \vec{r}$, 
and $\eta(\vec{r}'',t)$
is the time-dependent occupation variable of the well at position 
$\vec{r''}$ at time $t$;
$\eta(\vec{r''},t) = 1$ if the well is occupied by a 
monolayer particle and $\eta(\vec{r''},t) = 0$
if it is empty.

Finally, hard-core part of the interaction potential in Eq.(9) will be taken into account in the 
following way: we suppose that when the jump direction  is chosen, the particle 
attempts to jump into the target well. We stipulate, however, that 
the  jump can be only then  fulfilled, 
when at this time moment the target well is empty;
otherwise, the particle attempting this hop is repelled 
back to its position.

In such a picture of particles dynamics and interactions, 
which represents, in fact, 
the standard formulation of a hard-core lattice gas 
dynamics under long-range particle
interactions, 
the evolution of the local occupation variable
$\eta(\vec{r};t)$  can be described by an appropriate 
probabilistic 
generator $L\{\eta(\vec{r};t)\}$
 (see, e.g. \cite{leb,leba}). Here we will not go into the 
 details of rigorous probabilistic formulations, 
and will 
proceed
 by making a simplifying physical assumption that the realization average
of the product of the local occupation variables of different wells
factorize into the product of their average values, 
which corresponds to the assumption of local equilibrium.
It was shown recently in \cite{burb,burl} that such an assumption provides an adequate
description of particles dynamics in hard-core lattice gases and we thus expect that it
will be also a fair approximation for the system under study.
The assumption of the local equilibrium 
allows us to describe the system evolution in terms of local densities $\rho(\vec{r};t)$,
$\rho(\vec{r};t) = \overline{\eta(\vec{r};t)}$, which define the probability of having at time $t$
a particle in the well at position
$\vec{r}$. 
Consequently, instead of the probabilistic equations describing
evolution of $\eta(\vec{r};t)$, 
we will have to consider a deterministic integro-differential equation describing 
evolution of the 
the local densities  $\rho(\vec{r};t)$.
 In doing so,  we find then that the dynamics of $\rho(\vec{r};t)$
is governed by the following continuous-time equation
\begin{eqnarray}
\tau \;  \frac{\partial \rho(\vec{r};t)}{\partial t} \; = \; 
- \; \rho(\vec{r};t) \;  \sum_{\vec{r'}} 
 \overline{p(\vec{r}|\vec{r'})} 
\; \left(1 \; - \; \rho(\vec{r'};t)\right) \; 
+ 
&&
\nonumber \\  
+
\; \left(1 \; - \; \rho(\vec{r};t)\right) 
\;  \sum_{\vec{r'}}  \overline{p(\vec{r'}|\vec{r})} \; \rho(\vec{r'};t),
\end{eqnarray}
where the realization-average transition probabilities are given by
\begin{equation}
\overline{p(\vec{r}|\vec{r'})} \; = \; {\rm Z^{-1}} \; 
exp\left( - \frac{\beta U_{0}(T) \sigma^{6}}{2}   \left[\sum_{\vec{r''}}  
\frac{\rho(\vec{r''};t)}
{|\vec{r} - \vec{r''}|
^{6}} \; - \; \sum_{\vec{r''}} 
\frac{\rho(\vec{r''};t)}
{|\vec{r'} - \vec{r''}|^{6}}  \right] \right)
\end{equation}
Eq.(13) has a simple physical meaning - it describes the balance 
between the departures of a particle 
 from the well at position
$\vec{r}$ to any of the neighboring wells and the arrivals of particles 
from the neighboring wells to the well at
position $\vec{r}$. 
Particularly, the first term on the right-hand-side
of Eq.(13) describes all possible events in which  a particle, occupying at time $t$
the well at $\vec{r}$
 (the factor $\rho(\vec{r};t)$) may jump, at a rate $p(\vec{r}|\vec{r'})$ prescribed 
by the corresponding change in the energy of the monolayer,
to any of vacant (the factor $(1 - \rho(\vec{r'};t)$) adjacent
wells.
In a similar fashion,  the second term describes the corresponding (positive) 
contribution due to arrivals of particles from adjacent wells
to the well at position $\vec{r}$.

Eq.(13) has to be solved subject to the initial condition
\begin{equation}
\rho(\vec{r};0) \; = \; \left\{\begin{array}{lll}
0& \mathrm{for}&X \; > \; 0\\
\rho & \mathrm{for}&X \; \leq \; 0,\\
\end{array}
\right.
\end{equation}
where $\rho$ denotes the initial mean coverage 
(number of occupied wells per total number of wells in a unit
area) of the half-plane $X \leq 0$. Eqs.(13) and (15) represent 
the mathematical formulation of the problem
under study and allow for the computation of the monolayer edge
time
evolution.

\section{Approximations.}

One possible approach is to seek for an approximate  solution of Eqs.(13) and (15), 
turning to the continuous-space limit and
expanding the local densities into the Taylor series up to the second order 
in powers of $\sigma$ and  the exponentials in Eq.(14) up 
to the first order in the gradient terms.
In doing so, we obtain from our Eq.(13) the following 
continuous-space Fokker-Planck-type equation with non-local, configuration-dependent potential term
\begin{eqnarray}
\frac{\partial \rho(\vec{r};t)}{\partial t} \;  \; = 
\; D_{0}
  \; [ \triangle  \rho(\vec{r};t) \; - 
 \; \beta \; U_{0}(T) \; \nabla \; \{ \rho(\vec{r};t) \; \times
&&
\nonumber \\
\times \; 
( 1 \; - \; \rho(\vec{r};t)) \;  
 \int \; d\vec{r'}  
\;  \rho(\vec{r'};t) \; \nabla \;
\frac{1}{|\vec{r} - \vec{r'}|^{6}}\}]
\end{eqnarray}
Equation (16), which was 
rigorously derived by Giacomin and Lebowitz \cite{leba}
 for related lattice-gas model with Kac potentials, 
allows for an analytic, although rather complicated 
analysis (see \cite{leba} for more details).

Here we will pursue, however, more simple approach of \cite{bura,burc}, which
allows for a quite straightforward
 computation of the mean displacement of the monolayer edge
and of the edge tension directly from Eq.(13).  
Following \cite{bura,burc} we assume 
that for the long-range, but rapidly vanishing interaction
potentials as defined in Eq.(9), a hop of any monolayer particle being in the "bulk"
monolayer 
does not change the energy in Eq.(12). It means, in turn, that for such a particle
all hopping direction appear to be equally probable and the hopping events are constrained by 
the hard-core interactions only. This is certainly not so for the particles being directly at the edge
of the monolayer; these always have "free space" in front of them and the monolayer particles behind,
which will result in effective attraction of the edge particles to the "bulk" monolayer and, in
consequence, in asymmetric hopping probabilities: the edge particles will attempt preferentially
to hop towards the "bulk" monolayer. In reality, all particles of the monolayer
will experience a weak
attraction in the negative $X$-direction. 
Being zero for $X = - \infty$, the effective attractive force
will grow and reach its maximal value for $X = X(t)$, i.e. 
for the edge particles. 
Simplifying the actual picture to some extent, we will suppose here
that this "restoring" force is present only for the particles which are directly at the edge (see
also \cite{bura,burc}). 
This resembles, in a way, the model in \cite{joela,joelb,joelc},
which was concerned with the dynamics of the edge due to the edge tension only. 
In contrast to this model, however, the approach of \cite{bura,burc}
does not neglects the presence of the "bulk" monolayer phase; as we will see below
the hard-core interactions between the particles in the 
"bulk" monolayer are crucially important
and, particularly, are responsible for the $\sqrt{t}$-behavior 
in place of the linear in time dependence
predicted by Eq.(7).

In a more precise way, the main assumptions of the approach in \cite{bura,burc}
can be formulated as follows:\\

(a) Assume that at any time moment the monolayer
is homogeneous in the direction normal to the $X$-axis. Consequently, 
the energy $U(\vec{r};t)$ 
stays constant for any hop which does not change the particle position along the $X$-axis
and the probability for such  hops to take place 
is site-independent and is equal to $1/z$. 
This implies, in turn,
that the edge of the monolayer is sharp (straight line) 
and allows to reduce the problem
to the effectively one-dimensional model, 
in which the presence of the
second direction is accounted for only
through the renormalized diffusion coefficient and actual two-dimensional tension of the monolayer
boundary.\\ 
 
(b) In the general case the monolayer edge 
will change its position along the $X$-axis,
i.e. the monolayer will either contract or dilate in the $X$-direction, and thus
the density will be dependent on the $X$-coordinate. One may suppose, however, that 
the density distribution 
$\rho(X;t)$ will be a slowly varying, at a microscopic scale,  function of $X$, 
such that for weak long-range potentials in Eq.(9) and for 
$X$ which are strictly less than the instantaneous position of the monolayer edge $ X(t)$
the difference $\rm{U}
(X + \sigma;t)  -  \rm{U}
(X - \sigma;t) \ll 1/\beta$.  \\

(c)  Hopping probabilities
of particles being directly at 
the edge obey Eq.(14) and thus 
 depend implicitly
on both $X(t)$ and $\rho(X;t)$. 
On the other hand, one may expect that
after some transient period  of time 
(which will be not studied here),  the probability of making a  jump away
of the edge, i.e. $p(X(t)|X(t)+\sigma)$, and the probability of making a  jump
towards the "bulk" monolayer, i.e. $p(X(t) | X(t)-\sigma)$,  
approach some limiting values $p$ and $q$, which do not depend on $X(t)$ and $t$.
We note that this expectation is actually consistent 
with the solid-on-solid-model description of the liquid-vapour
 interface \cite{joela}. 
We will show below that this is actually the case
and stems from the stabilization of the density profile in the vicinity
of the moving edge.  

Finally,  $p$ and $q$ are to be determined  in a self-consistent way. 
To do this,
we will first solve the model described by (a) - (c) supposing that $p$ and $q$ are 
known,  fixed parameters, and  
calculate the time evolution of 
$X(t)$ and $\rho(X,t)$. Then, substituting the latter  into  Eq.(14),
we will obtain the closed-form transcendental equation, which determines 
the dependence of $p$ and $q$
on $U_{0}(T)$, $\rho$ and $\beta$. 

\section{Results.}

Let us consider now the approximate picture
of the monolayer evolution,
described  by (a) - (c), supposing first that $p$ and $q$ are some 
given parameters.
We notice that
such a picture has an evident interpretation 
within the framework of a lattice gas dynamics (Fig.3). 
 It defines, namely, the evolution
of a symmetric hard-core lattice gas,
which is initially placed, at mean density $\rho$,
at the sites $- \infty < X \leq 0$ of a one-dimensional 
infinite lattice of regular spacing $\sigma$. The particles are 
allowed to perform random hopping motion between the 
nearest-neighboring sites; the hopping motion is constrained
by hard-core interactions. All particles, excluding the rightmost 
particle of the gas phase, have symmetric hopping probabilities, i.e. for them an attempt
to hop to the right and an attempt to hop to the left occur with equal probability $m = 1/z \tau$.
In contrast, for the rightmost particle these probabilities are asymmetric; they
 are equal to $p$ and $q$ for jumps away of and towards the gas phase respectively.

Dynamics of the rightmost particle in such an asymmetric, (with respect to the density distribution
and the hopping probabilities of the rightmost particle), lattice-gas model
has been analysed recently in \cite{burl}. 
It was shown that for arbitrary values of the ratio $\mu = p/q$, ($0 \leq \mu \leq 1$),
the mean displacement of the rightmost particle  follows 
\begin{equation}
X(t) \; = \; A \; \sqrt{D_{0} t},  
\end{equation}
in which equation
 $D_{0}$ stands for the diffusion coefficient, $D_{0} = \sigma^{2}/z \tau$, 
and the parameter $A$ is determined
implicitly as the solution of the transcendental equation
\begin{equation} 
\frac{\sqrt{\pi} \; A}{2} \; \exp(A^{2}/4) \;  
[ 1 \; + 
\; \Phi(A/2)] \; = \; 
 \frac{\mu \; -  \; (1 -
\rho)}{1 - \mu}
\end{equation}
where $\Phi(x)$ denotes the probability integral. 
Besides, it was shown in \cite{burl} that at sufficiently large times
the density distribution past the rightmost particle
obeys
\begin{equation}
\rho(\lambda;t) \; = \; \frac{\rho}{1 + I(A)} \; \{1 \; + \; A^{2} \; \int^{\theta}_{0}
 dz \; exp(
- \frac{A^{2}}{2} (z^2 - 2 z))\},
\end{equation}
in which
\begin{equation}
I(A) \; = \; \sqrt{\frac{\pi}{2}} \; A \; exp(A^2/2) \; [1 \; + \; \Phi(A/\sqrt{2})],
\end{equation}
and $\theta$ is the scaled variable, $\theta = \lambda/A \sqrt{D_{0} t}$,   
where $\lambda$ stands for the relative distance from the rightmost particle, $\lambda =
X(t) - X$. In the limit $\lambda \ll \sqrt{D_{0} t}/A$,  Eq.(19) reduces to
\begin{equation}
\rho(\lambda;t) \; \approx \; (1 \; - \; \mu) \; 
[1 \; + \; \frac{A \; \lambda}{2
\sqrt{D_{0} t}} \; + \; ... \; ],
\end{equation}
while in the opposite regime, when
$\lambda \gg \sqrt{D_{0} t}/A$, the density  past the rightmost particle   approaches
the initial value $\rho$ exponentially fast. It is important to notice that Eq.(21) 
shows 
that the density  past the rightmost particle 
 is almost constant (and different from $\rho$) in a certain region whose size grows
in proportion to $X(t)$. We note finally that for the just described one-dimensional model
Eqs.(17) to (19) are exact, as it was shown subsequently by rigorous probabilistic analysis
in \cite{olla}.

Now, Eq.(18) predicts that four different regimes can take place, depending 
on the relation between $\mu$ and $\rho$.  
First, for $\mu < 1 - \rho$ the parameter $A$ is negative
and thus  the rightmost particle effectively
compresses the gas phase. When $\mu$ exactly equals $1 - \rho$, (which defines
the "yield" value of $\mu$ necessary to initiate further compression), the prefactor $A$ in Eq.(17)
is exactly equal to zero, 
$A = 0$. Thus  $X(t) = 0$ and the gas phase is stable. 
In this regime, however, despite the fact that $X(t) = 0$,
the rightmost particle still wanders randomly  around the equilibrium position 
 and its mean-square displacement $X^2(t)$ grows with time. In \cite{burl} it was shown that the growth
is
sub-diffusive and  $X^2(t) \sim (1 - \rho) t^{1/2}/\rho$.
Further on,   $A$ is positive and finite when $1 < \mu < 1 - \rho$ holds;
in this regime entropic effects overcome the pressure exerted by  the rightmost particle
and the gas slowly decompresses.  Finally,  
when $\mu = 1$, the rhs of Eq.(18) diverges, which means that $A$ is infinitely large
in the steady-state.  Actually, in this case $A$ shows a slow, logarithmic  growth with time.
At sufficiently large times,
\begin{equation}
A \; \approx \; \sqrt{2 \; ln(\frac{\rho^2 \omega t}{\pi})} 
\end{equation}

Turning now back to the system under study,  
we have to identify $\mu$ and to express it
through the
strength of interactions $U_{0}(T)$, initial coverage $\rho$ and the temperature $\beta^{-1}$. 
As we have
already mentioned, the asymmetry in the hopping probabilities
of the edge particles or, in other words, the
"edge" tension force 
  arises 
from the attraction of the edge particles to the "bulk" monolayer.  Thus,
since the interaction potential rapidly vanishes with the distance, we may expect that it is
mostly 
dominated by the density profile in the
vicinity of the moving edge, 
which is itself
dependent on the magnitude of the edge tension. 
Self-consistent choice of $\mu$ is thus
prescribed by Eq.(14), which relates the 
hopping probabilities $p$ and $q$ to the density profile in the
monolayer. Taking into account the result in 
Eq.(21), which defines the density profile in the vicinity of the moving edge,
 we find from Eq.(14) that the ratio $\mu = p/q$ obeys
\begin{equation}
\mu \; = \; exp( - \beta \sigma \gamma_{edge})
\end{equation}
In Eq.(23) the parameter $\gamma_{edge}$ stands for the edge tension,
\begin{equation}
\gamma_{edge} \; = \; (1 \; - \; \mu) \; \frac{U_{0}(T) \; \delta}{2},
\end{equation}
and $\delta$ can be thought of as the number of broken
cohesive bonds due to a hop away of the two-dimensional edge of the monolayer. Explicitly,
\begin{equation}
\delta \; = \; \sigma^{5} \; \sum_{\vec{r''}} \{\frac{1}{|\vec{r_{-}} - \vec{r''}|^6} \; - \; 
\frac{1}{|\vec{r_{+}} - \vec{r''}|^6}\},
\end{equation}
where the shortenings $\vec{r_{\pm}}$ stand for the vectors $(X(t) \pm \sigma,Y)$
and the sum extends over all lattice sites excluding $\vec{r''} = \vec{r_{\pm}}$. 
For the simplest case
of the square lattice, when $z = 4$, the parameter $\delta$ can be 
readily calculated in explicit form:
\begin{equation}
\delta \; = \; \sigma^{-1} [2 \; \sum_{j = 1}^{\infty} j^{-6} \; + 
\; \sum_{j = - \infty}^{\infty} (1 \; + \; j^{2})^{-3}] \; \approx \; 3.4 \; \sigma^{-1},
\end{equation}
which shows that in the presence of weak long-range attractive interactions $\delta$
only slightly exceeds $\delta_{c} = 3 \; \sigma^{-1}$ - the result
 which we would obtain 
in the extreme case of nearest-neighbor (Ising-type) attractions.

Therefore, we have that the
 mean displacement  of the monolayer edge
obeys Eq.(17), i.e. $X(t) = A \sqrt{D_{0} t}$, which is consistent with the behavior predicted
in \cite{bura,burc} for spreading ($A \geq 0$) precursor films.
In our case of a semi-infinite monolayer,  the parameter $A$ in this growth law is defined 
implicitly  through  Eqs.(18), (24) and (25), which allow to interpret it in terms of 
the strength of the PP interactions, 
initial mean coverage $\rho$
 and the temperature. 
In Fig.4 we present numerical solution of these equations and 
plot $A$ versus the dimensionless parameter $\epsilon = \beta U_{0}(T) \sigma \delta/2$, which appears
to be the most significant
 critical parameter of the model. The solid curves in Fig.4 define
the function $A(\epsilon)$
for four different initial densities.  These curves
demonstrate that the evolution of the monolayer is, in general, very sensitive
to the value of $\epsilon$ and it may proceed rather differently, 
depending on the relation between this parameter
and $\rho$.

We continue with some  analytical  
estimates of the $\epsilon$-dependence of the prefactor $A$, where we can specify four different
regimes. \\

I. When  $\epsilon$ 
belongs to a finite interval 
$0 \leq \epsilon \leq 1$ (high temperatures or low PP attraction), Eqs.(22) and (23)
possess only one trivial solution $\mu = 1$, which means that in this range
of parameters the "edge" tension is exactly equal to 
zero and the monolayer thus spreads as  
a surface gas.  The edge in this regime advances a
bit faster that pure $\sqrt{t}$-law and follows $X(t) \sim (t \; ln(t))^{1/2}$.
The particle density past the edge is almost zero within an extended interval, 
which grows in proportion to 
$X(t)$.\\

II. In the range  $1 < \epsilon < \epsilon_{c}$, where 
\begin{equation}
\epsilon_{c} \; = \; - \; \frac{ln(1 - \rho)}{\rho},
\end{equation}
the parameter $A$ 
is finite and positive. Therefore, in this regime the monolayer also wets the substrate 
and the edge displaces in proportion to $\sqrt{t}$. It is easy to check that in this regime the
edge tension $\gamma_{edge}$ is positive and vanishes as 
\begin{equation}
\gamma_{edge} \; \sim \; (T_{b} - T)
\end{equation}
when the temperature $T$ approaches the value $T_{b}$. The critical temperature $T_{b}$ is implicitly
defined by the condition $\epsilon = 1$, which can be rewritten as 
\begin{equation}
T_{b} \; = \; \frac{U_{0}(T_{b}) \; \sigma \; \delta}{2}
\end{equation}
Now, since the edge tension is positive below the $T_{b}$ and is exactly zero above the $T_{b}$, it
seems natural to identify this regime as the regime of liquid-like spreading and, correspondingly, the
temperature $T_{b}$ - as the temperature of the surface gas-liquid transition or, in other words, as the
boiling temperature of the monolayer on solid surface. We note finally that the parameter $A$ diverges
in the limit $T \to T_{b}$, (when $\epsilon \to 1$),  
\begin{equation}
A \; \sim \; \sqrt{ln(\frac{\rho}{\epsilon - 1})}
\end{equation}
Within the opposite limit $\epsilon \to \epsilon_{c}$, the parameter $A$ vanishes as
\begin{equation}
A \; \sim \; \frac{(1 - \rho) (\epsilon_{c} - \epsilon)}{1 - (1 - \rho) \epsilon_{c}}
\end{equation}
In this liquid-like spreading regime, 
the particle density past the edge is nearly constant within a 
region of size $X(t)$ and is lower than the unperturbed density $\rho$ 
in the bulk monolayer.\\

III. At the point $\epsilon = \epsilon_{c}$ the monolayer partially wets the substrate; 
$\rho(\lambda;t) = \rho$ and the prefactor
$A$ is exactly zero. We thus denote this point as the point of the wetting/dewetting transition.  The
corresponding critical temperature $T_{w/dw}$ 
is defined by Eq.(27), which gives, explicitly,
\begin{equation}
T_{w/dw} \; = \; \frac{U_{0}(T_{w/dw}) \; \sigma \; \delta}{2 \epsilon_{c}}
\end{equation}
Thus, this critical temperature appears 
to depend on the monolayer coverage $\rho$. We note now that two
critical temperatures are simply related to each other. When $U_{0}(T)$ is independent of $T$, like
it is in the case of the London-van der Waals interactions,  we find the following relation
\begin{equation}
T_{w/dw} \; = \; \frac{T_{b}}{\epsilon_{c}}
\end{equation}
For the Keesom-van der Waals interactions, when $U_{0}(T) \sim 1/T$, we find instead of Eq.(33),
\begin{equation}
T_{w/dw} \; = \; \frac{T_{b}}{\sqrt{\epsilon_{c}}}
\end{equation}
For the monolayer, the wetting/dewetting transition point $\epsilon_{c} \to \infty$ when $\rho \to 1$;
consequently, the critical temperature of the wetting/dewetting transition $T_{w/dw} \to 0$ in this
limit. Within the opposite limit, i.e. when $\rho \to 0$, $T_{w/dw} \to  T_{b}$.\\

IV. Finally, for $\epsilon > \epsilon_{c}$,  
which corresponds to the limit of either low temperatures or
strong particle-particle attractions, the parameter $A$ is negative, 
$A < 0$, and thus the presence of a monolayer with a given coverage $\rho$ 
on the solid surface is energetically
non-favorable; consequently, it
dewets from the substrate. 
The $\epsilon$-dependence of the parameter $A$ appears to be very weak in
this regime (see Fig.4), which means that compressibility of the monolayer is very low, being strongly
limited by the process of diffusive squeezing out of "voids"  at progressively larger
and larger scales. The density before the retracting edge is higher than the mean value $\rho$
in an extended region which grows in proportion to $X(t)$ (an analog of the rim in the hydrodynamic dewetting 
\cite{broch,pgg3,red,shana}).

We hasten to remark, however, that the predicted weak $\epsilon$-dependence of the parameter $A$
 concerns only the 
situation in which thickening of the monolayer
is not allowed and in which  
the particle motion can be viewed as an activated hopping motion between the wells created by the
 particle-solid interactions; particle-particle interactions are assumed
to be small compared to the particle-solid interactions, such that they can be treated only as a
small perturbation.   
For liquids in confined geometries, 
where the geometrical constraints themselves do not allow for the
thickening of the monolayer, the process of 
 squeezing of voids out of
the "bulk" monolayer will be the
only mechanism of the dewetting process. 
However, for sufficiently strong particle-particle attractions, 
comparable to the diffusive barrier $V_{l}$, our approximate description of particles dynamics
is not justified; 
consideration of the edge tension as the only driving force will not be appropriate
either.

For monolayers on open solid substrates, thickening of the monolayer by 
forming progressively 
higher and higher "towers"
 \cite{pgg1}, represents an additional mechanism of the dewetting process,
which may be, under certain conditions, 
more 
efficient than diffusive squeezing of voids.
One can thus expect that for sufficiently strong particle-particle interactions
 the dewetting will be facilitated by thickening of the film, 
resulting in more pronounced $\epsilon$-dependence 
of the parameter $A$. 
As found in 
\cite{pgg1}, in this regime the mean displacement of the edge still 
follows the $\sqrt{t}$-dependence,
which means that $A$ remains finite in the limit $t \to \infty$. 
We may, however, only
speculate about the $\epsilon$-dependence 
of the parameter $A$ for such a process, since such a 
possibility is not included into the model. We sketch  
in Fig.4  a hypotetical behavior 
in this regime (the curve given by  squares). 
Lastly, when $U_{0}(T)$ becomes comparable to 
 the
adsorption barrier $U_{d}$, one may expect 
transition to the hydrodynamic dewetting, when the monolayer
tends to form a macroscopically large
 droplet. 
This regime was examined (starting from sufficiently
 thick initial
films, however) in \cite{broch,pgg3,red,shana} 
and it is known that the edge in this regime displaces
at a constant velocity, which means that here $A$ does not tend 
to a constant value as $t \to \infty$,
but rather increases indefinitely as time evolves, 
$A \sim - t^{1/2}$. 
In Fig.4 we 
mark the transition from the monolayer-dewetting regime to the hydrodynamic
dewetting regime by the line of crosses.
 We can not, of course, identify precisely
the  value of the parameter
$\epsilon$ at which such a transition takes place; this calculation
requires, again, elaboration of
the model allowing for the thickening of the monolayer.

We finally comment that our results interpret the 
notion of the two-dimensional volatility,
which is commonly used in experimental 
literature on spreading of molecularly thin films, in terms of 
the parameters of the particle-particle
interactions, particle density  and the temperature. 
It is precisely the relation between the value of the
parameter $\epsilon$,
which is the measure of the particle-particle 
cohesive interactions, and the critical value
$\epsilon_{c}$, which shows whether having a monolayer on the solid substrate
is energetically favorable or not. Consequently, we may expect that liquids
with $\epsilon \geq \epsilon_{c}$ are not volatile in two dimensions, while
liquids with $\epsilon > \epsilon_{c}$ are. 

\section{Conclusions.}

To summarize, we have presented a microscopic dynamical description of the time
evolution of a monolayer on solid surface. The monolayer was assumed
to be created in initially  non-equilibrium
configuration, in which it covers only one half of the solid surface,
and then was allowed to evolve in time due to particles random motion.
Particles hopping motion was determined as the Kawasaki particle-void exchange dynamics 
in presence of long-range  particle-particle attractions \cite{leba}. 
Here we have focused exclusively 
on a two-dimensional behavior, assuming that particle evaporation from the substrate is absent
and thickening of the monolayer is forbidden. 
We have 
shown that in such a situation the behavior of the monolayer is
very sensitive to physical conditions and parameters of the particle-particle attractions.
Depending on the strength of the latter, the monolayer can show different kinetic behavior;
it can wet, partially wet or spontaneously dewet from the substrate. 
More precisely, our results can be summarized as follows. We find that the mean displacement
of the monolayer edge $X(t)$ evolves as $X(t) = A (D_{0} t)^{1/2}$, where
$D_{0}$ is the bare diffusion coefficient describing dynamics of an isolated
particle on top of solid surface, while $A$ is some parameter dependent on
 the strength of liquid-liquid attractions $U_{0}(T)$ and temperature $T$. 
At sufficiently high temperatures, such that $T \geq T_{b}$, 
the parameter $A$ is greater than zero and shows a slow growth with time,
$A \sim (ln(t))^{1/2}$. We identify this regime as "surface-gas" spreading, since 
the tension of the monolayer edge
appears to be equal to zero. Actually, in this regime the density past the edge 
is almost zero in an extended interval, whose size grows in proportion to $t^{1/2}$.
Next, 
in the range of temperatures
$T_{b} > T > T_{w/dw}$ the parameter $A$ is also positive
but tends to a certain constant value as $t \to \infty$. 
The edge tension in this regime is positive
and
we thus call it as the regime of "liquid-like"
spreading. The density past the edge is constant within the interval of size $X(t)$
and is less than the density in the bulk monolayer.
Thus in both regimes the monolayer expands
and wets the substrate.
Further on, at $T = T_{w/dw}$ 
 the parameter $A = 0$, i.e. 
 there is no regular dependence 
of the displacement of the edge on time and the
 monolayer remains in its initial configuration. 
  Finally,
below the temperature of the wetting-dewetting
 transition, 
$T_{w/dw}$, the parameter $A$ is
 constant and negative, i.e. the monolayer
contracts by squeezing out the "voids" and dewets
 from the solid surface.

\vskip 10mm

{\bf \Large Acknowledgments} \nopagebreak \\ \nopagebreak
\vskip 0.2cm

The authors acknowledge helpful discussions with S.F. Burlatsky, 
T. Blake, J. Lyklema, J.L. Lebowitz, J. Ralston and E. Rapha\"el.  
We also thank E. Tosatti for bringing our attention to 
Ref.37.
Financial support from the FNRS, the COST Project D5/0003/95 and 
the EC Human and Capital Mobility Program
CHRX-CT94-0448-3 is gratefully acknowledged.


\vskip 0.3cm

\begin{thebibliography}{99}

\vskip 0.2cm
%
\bibitem{hardy} W.Hardy,  Phil. Mag. 38 (1919) 49
%
\bibitem{bang} D.Bangham and S.Saweris,  Trans. Faraday Soc.  33 (1938) 554 
%
\bibitem{bas}  W.Bascom, R.Cottington and C.Singleterry, in: Contact Angle, Wettability and
Adhesion, ed.: F.M.Fowkes, Advances in Chemistry, Vol.  43, American Chemical Society,
Washington DC, 1964, p.355
%
\bibitem{caz0} A.M.Cazabat, Contemp. Phys.  28 (1987)  347 
%
\bibitem{pgg0}   P.G.de Gennes, Rev. Mod. Phys. 57  (1985) 827 
%
\bibitem{der0}   B.V. Derjaguin and N.V. Churaev, Wetting Films, Nauka, Moscow, 1984;
B.V. Derjaguin, N.V. Churaev and V.M. Muller,  Surface Forces, Consultant Bureau, New York,
1987
%
\bibitem{lyk} J.Lyklema, Fundamentals of Interface and Colloid Science, Vol.1, 
Academic Press, London, 1991
%
\bibitem{caze} A.M.Cazabat and M.A.Cohen Stuart, J. Phys. Chem. 90 (1986) 5845
%
\bibitem{joa} J.F.Joanny and P.G.de Gennes, J. Phys. (Paris) 47 (1986) 121
%
\bibitem{joab} J.F.Joanny and P.G.de Gennes, C. R. Acad. Sci. 299 II (1984) 279; 605
%
\bibitem{ruck1} E.Ruckenstein, in: Metal-Support Interactions in Catalysis, Sintering and
Redispersion, eds.: S.A.Stevenson, J.A.Dumesic, R.T.K.Baker and E.Ruckenstein, Van
Nostrand-Reinhold, New York, 1987
%
\bibitem{ruck2} E.Ruckenstein, J. Colloid Interface Sci. 179 (1996) 136
%
\bibitem{cazy} A.M.Cazabat et al., Adv. Colloid Interface Sci. 48 (1994) 1
%
\bibitem{caza}  F.Heslot, N.Fraysse and A.M.Cazabat, Nature (London) 338 (1989) 640
%
\bibitem{cazb} F.Heslot, A.M.Cazabat and P.Levinson, Phys. Rev. Lett. 62 (1989) 1286
%
\bibitem{cazc} F.Heslot, A.M.Cazabat and N.Fraysse, J. Phys. Cond. Mat.  1 (1989) 5793
%
\bibitem{cazd}  J.De Coninck, N.Fraysse, M.P.Valignat and A.M.Cazabat, Langmuir 9 (1993) 
1906
%
\bibitem{cazf}  A.M.Cazabat, J.De Coninck, S.Hoorelbeke, M.P.Valignat and S.Villette, Phys. Rev. E
49 (1994) 4149 
%
\bibitem{pgg2} P.G.de Gennes and A.M.Cazabat,  C. R. Acad. Sci. 
Paris 310  (1990) 1601
%
\bibitem{vil} S.Villette, J.De Coninck, F.Louche, 
A.M.Cazabat and M.P.Valignat, to be published
%
\bibitem{joela}  D.B.Abraham, P.Collet, J.De Coninck and F.Dunlop, 
Phys. Rev. Lett. 65 (1990) 195
%
\bibitem{joelb}  D.B.Abraham, P.Collet, J.De Coninck and F.Dunlop,
 J. Stat. Phys.  61 (1990) 509
%
\bibitem{joelc} J.De Coninck, F.Dunlop and F.Menu, Phys. Rev. E  47 (1993) 1820
%
\bibitem{lif} I.M.Lifschitz, Sov. Phys. JETP 15 (1962) 939
%
\bibitem{joelx} J.De Coninck et al., work in progress
%
\bibitem{bura} S.F.Burlatsky, G.Oshanin,  A.M.Cazabat and   M.Moreau, Phys.  Rev. Lett. 
 76 (1996) 86
%
\bibitem{burc}  S.F.Burlatsky, G.Oshanin, A.M.Cazabat,  M.Moreau 
and W.P.Reinhardt, Phys. Rev. E 54 (1996) 3832
%
\bibitem{burd}  S.F.Burlatsky, A.M.Cazabat, M.Moreau,  
G.Oshanin and S.Villette, in: Instabilities and Non-Equilibrium Structures VI, ed. E.Tirapegui, Kluwer
Academic Publ., Dordrecht, to appear; preprint cond-mat/9607143
%
\bibitem{ort} J.De Coninck, U.D'Ortona, J.Koplik and J.R.Banavar, Phys. Rev. Lett. 
74 (1995) 928
%
\bibitem{col} J.De Coninck, Colloids and Surfaces 114 (1996) 155 
%
\bibitem{ala} O.Venl\"ainen, T.Ala-Nissila and K.Kaski, Physica A
  210 (1994) 362
%
\bibitem{pgg1} D.Ausserr\'e, F.Brochard-Wyart and P.G.de Gennes,  C. R.  
Acad. Sci. Paris 320 (1995) 131
%
\bibitem{broch} F.Brochard-Wyart and P.G.de Gennes,  Adv. Colloid Interface Sci.  39 (1992) 
1
%
\bibitem{pgg3}  P.G.de Gennes, in: Physics of Amphiphilic Layers, Vol.34, Springer-Verlag,
Berlin, 1987, p.64
%
\bibitem{red} C.Redon,  F.Brochard and F.Rondelez, Phys. Rev. Lett.  66 (1991) 715 
%
\bibitem{shana} A.Carr\'e and M.E.R.Shanahan, Langmuir  11 (1995) 3572 
%
\bibitem{tos} B.N.J.Persson and E.Tosatti, Phys. Rev. B  50 (1994) 5590
%
\bibitem{blakea} T.D.Blake and J.M.Hayes, J. Colloid Interface Sci.  30 (1969) 421%

\bibitem{blakeb} T.D.Blake, Dynamic Contact Angles and Wetting Kinetics, in: Wettability, ed.: J.C.Berg,
Marcel Dekker, New York, 1993
%
\bibitem{mich} T.D.Blake, A.Clarke, J.De Coninck and M.J.de Ruijter, Langmuir  13 (1997) 2164
%
\bibitem{adam} A.W.Adamson, Physical Chemistry of Surfaces, Wiley-Interscience Publ., New-York,
1990
%
\bibitem{clark} A.Clark, The Theory of Adsorption and Catalysis, Academic Press, New York, 1970, Ch.2
%
\bibitem{eyr} J.J.McAlpin and R.A.Pierotti, J. Chem. Phys.  41 (1964) 68; 42 (1965) 1842
%
\bibitem{dev} A.F.Devonshire, Proc. Roy. Soc. (London), Ser.A 163 (1937) 132
%
\bibitem{dif} Diffusion at Interfaces: Microscopic Concepts, eds.: M. Grunze, 
H.J. Kreuzer and J.J. Weimer,
Springer Series in Surface Sciences, Vol.12, Springer-Verlag, Berlin, 1988
%
\bibitem{zang} A.Zangwill, Physics at Surfaces, Cambridge University Press, Cambridge, 1988
%
\bibitem{maz} S.Ramaswamy and G.Mazenko, Phys. Rev. A  26 (1982)  1735
%
\bibitem{gor} Z.W.Gortel and L.A.Turski, Phys. Rev. B 45 (1992) 9389
%
\bibitem{pom} U.Frisch, B.Hasslacher and Y.Pomeau, Phys. Rev. Lett.  56 (1986) 1505
%
\bibitem{gom} M.C.Tringides and R.Gomer, Surf. Sci.  265 (1992) 283
%
\bibitem{leb} J.L.Lebowitz, E.Orlandi and E.Presutti, J. Stat. Phys.  63 (1991) 933
%
\bibitem{leba} G.Giacomin and J.L.Lebowitz, Phys. Rev. Lett. 76 (1996) 1094;
  J. Stat. Phys. 87 (1997) 37
%
\bibitem{tur} M.A.Zaluska-Kotur and L.A.Turski, Phys. Rev. B  50 (1994) 16102
%
\bibitem{burb}  S.F.Burlatsky, G.Oshanin, M.Moreau and W.P.Reinhardt, Phys. Rev. 
E  54 (1996) 3165
%
\bibitem{burl}  G.Oshanin, J.De Coninck, M.Moreau and S.F.Burlatsky, Dynamics of 
the shock front
propagation in a one-dimensional hard-core lattice gas, J. Stat. Phys., to appear
%
\bibitem{olla} C.Landim, S.Olla and S.B.Volchan, Driven tracer 
particle in one dimensional symmetric simple
exclusion, Commun. Math. Phys., to appear
%
\end{thebibliography}
\end{document}